 \documentclass[preprint]{aastex}

\newcommand{\degrees}{$^{\circ}$}

\shorttitle{Stacking}
\shortauthors{Kurczynski et al.}

\usepackage{graphicx}

\DeclareGraphicsExtensions{.pdf,.png,.jpg,.eps}

\usepackage{amssymb, amsmath}

\begin{document}


\title{A Simultaneous Stacking and Deblending Algorithm for Astronomical Images}



\author{Peter Kurczynski}
\and
\author{Eric Gawiser}

\affil{Department of Physics and Astronomy, Rutgers University,
    Piscataway, NJ 08854}




\begin{abstract}

Stacking analysis is a means of detecting faint sources using a priori position information to estimate an aggregate signal from individually undetected objects.  Confusion severely limits the effectiveness of stacking in deep surveys with limited angular resolution, particularly at far infrared to submillimeter wavelengths, and causes a bias in stacking results.  Deblending corrects measured fluxes for confusion from adjacent sources; however, we find that standard deblending methods only reduce the bias by roughly a factor of two while tripling the variance.  We present an improved algorithm for simultaneous stacking and deblending that greatly reduces bias in the flux estimate with nearly minimum variance.  When confusion from neighboring sources is the dominant error, our method improves upon RMS error by at least a factor of three and as much as an order of magnitude compared to other algorithms.  This improvement will be useful for Herschel and other telescopes working in a source confused, low signal to noise regime.

\end{abstract}





\section{Introduction}

At every wavelength of astronomy, there is a need to extract faint signals buried in noisy images.  For example, a certain class of objects may be lurking undetected below the noise threshold of a data set at a particular wavelength.  However, if accurate positions for the objects are known a priori from detections at other wavelengths, then this prior information may be exploited to break through the noise floor and achieve a detection, at least in a statistical sense, for the objects in question.  

Stacking is an effective analysis technique in cases where noisy images at one wavelength are complemented by catalogs of known objects that have been detected in the same region of sky at other wavelengths.  The noisy data at known object positions are averaged, and the resulting aggregate flux measurement has an enhanced signal to noise ratio due to averaging, which reduces the noise by a factor $1/\sqrt{R}$, where $R$ is the number of target objects in the stack.

As an early application of this method in x-ray astronomy, stacking analysis was used to study x-ray emission from stars \citep{1985ApJ...289..279C}; subsequently, stacking became a standard x-ray analysis technique that, among others, has been applied to normal galaxies \citep{2001AJ....122....1B}, Lyman Break Galaxies (LBGs) \citep{2001ApJ...558L...5B} and radio sources \citep{2003MNRAS.345..939G}.

At optical to near infrared wavelengths stacking has also become widely adopted, and has been used for example to study galactic halos \citep{Zibetti2004} and intergalactic stars \citep{Zibetti2005}, as well as the cosmic infrared background \citep{2006A&A...451..417D}, star forming galaxies \citep{2004ApJ...610..745L, 2007ApJ...671..278G} and Extremely Red Objects (EROs) \citep{1997AJ....113..474H}.  At submillimeter wavelengths, stacking has been employed to study the submillimeter background \citep{2000MNRAS.318..535P, 2004ApJS..154..118S, 2006ApJ...647...74W, 2006ApJ...644..769D, 2009arXiv0904.1205M, 2009arXiv0904.0028G}, optical/IR color selected galaxies \citep{2004ApJ...605..645W, 2005ApJ...631L..13D, 2007MNRAS.381.1154T, 2009MNRAS.394....3D, 2009arXiv0904.0028G}, and radio detected galaxies \citep{2008MNRAS.385.2225S}.  Similarly, at radio wavelengths stacking has been widely adopted, for example to study LBGs \citep{2007ApJ...660L..77I, 2008ApJ...689..883C}, star forming BzK galaxies \citep{2005ApJ...631L..13D, 2007MNRAS.381.1154T, 2009MNRAS.394....3D}, Distant Red Galaxies (DRGs) \citep{2007ApJ...660L..77I}, EROs \citep{2009MNRAS.394....3D} and quasars \citep{2007ApJ...654...99W}.

It has been widely recognized that confusion noise limits astronomical measurements particularly at long wavelengths \citep{1974ApJ...188..279C, 2001AJ....121.1207H} and confusion poses special complications for stacking analyses.  When multiple sources crowd a resolution element of the data, then blending of sources must be accounted for in the stacking procedure.  Blending is a particular problem in the submillimeter regime, where at present images suffer from poor resolution compared to many other astronomical observations.  Methods have been used for deblending in submillimeter data, however a systematic assessment of the effectiveness of these techniques has not been previously reported.

This paper addresses the problem of deblending the various unresolved sources of submillimeter emission in order to effectively stack objects in the confusion dominated, low signal to noise, regime.  We present an algorithm that accomplishes simultaneous stacking and deblending of these data and demonstrate its effectiveness through Monte Carlo simulation.  We also demonstrate that the standard approaches to deblending are in fact subject to substantial bias and scatter when applied to real data.

\section{Data}

The stacking methodology presented here arose out of a study of the submillimeter emission of star forming galaxies as detected by the Large Apex BOlometer CAmera (LABOCA) \citep{2009A&A...497..945S} in the LABOCA Extended Chandra Deep Field South Submillimeter Survey (LESS) \citep{2009arXiv0910.2821W}.  The LESS survey covers the 30$' \times$30$'$ Extended Chandra Deep Field South (ECDF-S) at 870 $\mu$m to a noise level of $\sigma_{870 \mu m} \approx$ 1.2 mJy beam$^{-1}$.  The LESS catalog contains 126 individually detected submillimeter sources \citep{2009arXiv0910.2821W}; however, buried within the noise of these data are signatures from a multitude of faint, extragalactic objects that can only be detected in the aggregate via stacking analyses \citep{2009arXiv0904.0028G}.  

The MUltiwavelength Survey by Yale Chile (MUSYC) \citep{2006ApJS..162....1G} $K$ band source catalog \citep{2009ApJS..183..295T} contains more than 8,000 objects with $K$ magnitude less than 20 in the ECDF-S.  Positions of these $K<20$, predominantly extragalactic objects constitute the prior information used in stacking the submillimeter data.  Optical color selection techniques permit the classification of various subsets of these data into different high redshift galaxy types.  Stacking these subsets enables investigation of their submillimeter properties, which is the scientific motivation for these investigations.

\begin{figure}[h]
\begin{center}

\includegraphics[scale=0.40]{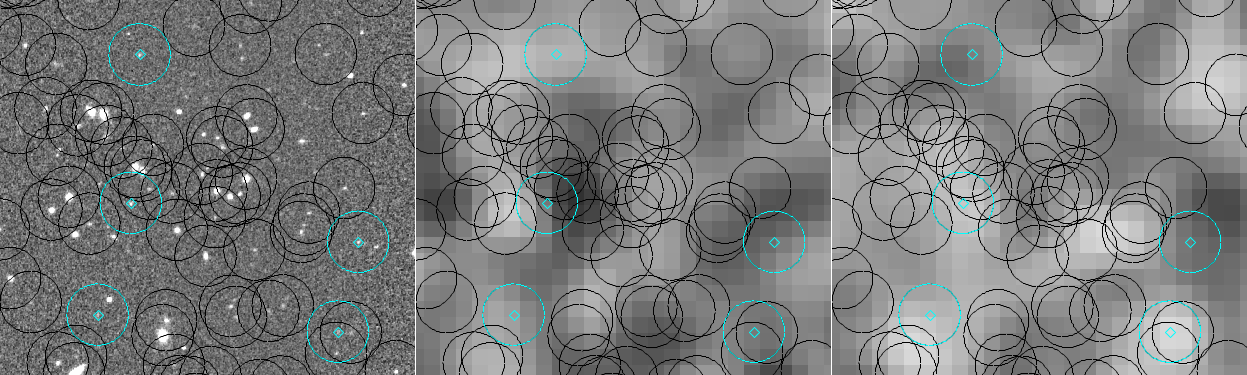}

\vspace{-0.5cm}
\end{center}
\caption{\footnotesize
{Left panel:  a subregion of the ECDF-S in K band.  Middle panel:  LESS 870 $\mu$m data for the same region.  Right panel:  an artificial data and noise realization from simulations.  Black circles in each image correspond to MUSYC $K<20$ object positions; radii are the LABOCA 1$\sigma$ beamwidth.  Cyan circles and diamonds indicate target positions for stacking analysis.}}

\label{FIG:ECDFSData}
\end{figure}

An example of the imaging data from the ECDF-S is shown in Figure \ref{FIG:ECDFSData}.  This figure shows co-aligned subregions of the ECDF-S in K band (left panel) and at 870 $\mu$m (middle panel).  Circles around K band sources are set to the LABOCA instrument beamwidth, 27.6$''$ FWHM ($\approx$ 12$''$ 1$\sigma$).  Diamonds indicate the positions of star forming, sBzK galaxies, as determined from optical-NIR colors \citep{2008ApJ...681.1099B}.  It is clear from the figures that these galaxies are interspersed with a large number of other $K$ band sources.  Typically each stacking target position has between one and ten other objects within one LABOCA beam diameter.  These sources are found to consist predominantly of other galaxy types that may themselves emit confusing submillimeter flux that blends with the flux from the target objects.  

The beam smoothed, LABOCA residual map, consisting of data with individual sources subtracted, was used as the basis of investigations presented here.  The map had a Gaussian flux density distribution, with a positive excess due to faint undetected sources \citep{2009arXiv0910.2821W};  because the average flux had already been subtracted from this map as part of data reduction, source populations that were not spatially correlated with the stacking target population would not have contributed to the average stacked flux, although they could have caused scatter in the flux estimate.

\section{Stacking Methodology}
\label{MethodologySection}

Estimating the aggregate flux of a set of $R$ individually undetected (i.e. detected with very low signal to noise ratio) sources with known, prior positions, can be done by averaging the flux values at the position of each undetected source.  This method is known as stacking.  With a sufficiently large number of source positions, a statistically significant aggregate signal, or stacking detection, can be obtained since the error of the weighted average diminishes by $1/\sqrt{R}$.\footnote{assuming Gaussian statistical errors of the individual flux measurements.}  However, known prior information typically includes a catalog of $\Lambda$ total sources, that consist of $R$ desired stacking targets, and an additional $F \equiv \Lambda - R$ non-target objects whose flux contributions must be assumed to confuse and blend with the target objects.  The aim of this paper is to demonstrate an effective method of extracting a stacking detection in the presence of these confused sources.  Following the approach introduced by \citet{2009arXiv0904.0028G}, we consider blending of flux contributions from all detected source types, including target and non-target populations, as opposed to considering only blending between target objects.

If the intrinsic flux due to a source at pixel $i$ is given by $I_i$, and the statistical measurement error is given by $\sigma_i$, then the inverse variance weighted average flux estimate is found from elementary statistics by summing over the pixels corresponding to all target sources in the stack and is given by
\begin{equation}
\left<I \right> = \frac{\sum \frac{I_i}{\sigma_i^2}}{\sum \frac{1}{\sigma_i^2}}
\label{EQ:wtavg}
\end{equation}
The corresponding error of the estimate, $\sigma_{\left<I \right>}$, is found from
\begin{equation}
\frac{1}{\sigma_{\left<I \right>}^2} = \sum \frac{1}{\sigma_i^2}
\end{equation}
The resulting signal to noise ratio of the stacking estimate is given by $SNR \equiv \left< I \right> / \sigma_{\left< I \right>}$.

As an estimator, the median is more robust to outliers, but has some ambiguity in interpretation of the errors and the detection significance.  Because this paper is concerned mainly with the effects of the random statistical errors on the significance of stacking measurements, the inverse variance weighted average is used herein, unless otherwise indicated.  

The high density of sources indicated in Figure \ref{FIG:ECDFSData} means that individual flux measurements, denoted as $f_i$ below,  include blending of intrinsic flux, $I_i$, from nearby and overlapping sources, and therefore individual flux measurements must be deblended to obtain the true source fluxes.


\subsection{Standard Deblending}

\begin{figure}[h]
\begin{center}

\includegraphics[scale=0.40]{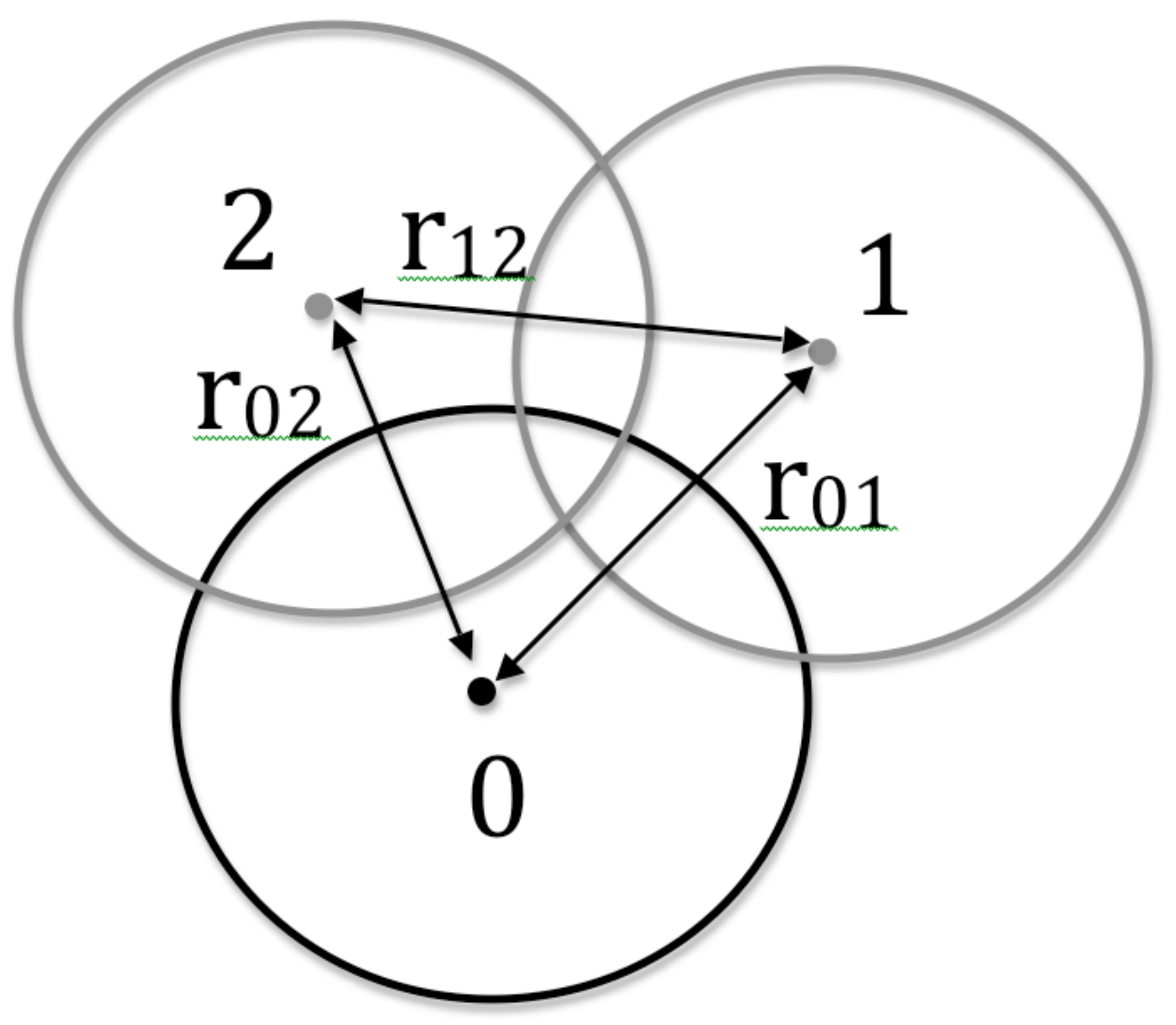}

\vspace{-0.5cm}
\end{center}
\caption{\footnotesize
{Illustration of confused sources.  Known object positions are represented by dots, and the submillimeter beamwidth is indicated by circles surrounding each object.  The stacking target, labeled 0, is found near adjacent, non-target sources, labeled 1, 2, with separations as indicated.}}

\label{FIG:Diagram}
\end{figure}

Deblending of flux from adjacent sources is illustrated in Figure \ref{FIG:Diagram}.  The measured fluxes, $f_0, f_1, f_2$ at the target source position and two adjacent source positions are related to the true, intrinsic source fluxes, $I_0, I_1, I_2$ in Equations \ref{EQ:deblend1}-\ref{EQ:deblend3} below.  This formalism assumes that source locations are at pixel center; in submillimeter data pixel stacking is normally performed on a beam-smoothed map, so $f_i$ is naturally interpreted as a pixel flux.  Gaussian beamwidth factors that include the separation $r_{ij}$ between sources at positions $i$ and $j$ and the instrument beamwidth\footnote{In far IR and submillimeter astronomy, the FWHM is the preferred measure of instrument resolution, and is related to the instrument beamwidth defined here according to $\Sigma$ = FWHM / 2.35}, $\Sigma$, are defined as $\alpha_{ij} \equiv e^{-r_{ij}^2/2 \Sigma^2}$, where $\alpha_{ii} = 1$, or $r_{ii} = 0$.  Noise terms $n_i$ are included below.
\begin{equation}
n_0 + \alpha_{00} I_0 + \alpha_{01}I_1 + \alpha_{02}I_2 = f_0
\label{EQ:deblend1}
\end{equation}
\begin{equation}
n_1 + \alpha_{10} I_0 + \alpha_{11}I_1 + \alpha_{12}I_2 = f_1
\label{EQ:deblend2}
\end{equation}
\begin{equation}
n_2 + \alpha_{20} I_0 + \alpha_{21}I_1 + \alpha_{22}I_2 = f_2
\label{EQ:deblend3}
\end{equation}
In the absence of noise, $n_i \rightarrow 0$, and this 3x3 system is exactly solvable for the intrinsic source fluxes.   It is easily generalized to include the effects of an arbitrary number, $N$, of neighbors.  This method of deblending is referred to as standard deblending in the simulations described in Section \ref{SimulationsSection}.  When noise is present, it propagates into the results of matrix inversion for $I_0$, $I_1$, and $I_2$.  An even more general set of next-nearest neighbor deblend equations were implemented in \citet{2009arXiv0904.0028G}; their approach applied these equations iteratively for each neighbor's neighbors with the process terminating once no further neighbors were found within a beamwidth.  

\subsection{Co-deblending}
\label{CoDeblendingSection}

The method outlined above can be made more robust in the presence of noise by making an assumption that all target objects have a uniform, a priori unknown, intrinsic flux, and all non-target objects have a different, uniform intrinsic flux.  This assumption is reasonable because the nature of stacking is to sacrifice knowledge of individual sources in favor of a more sensitive, aggregate measure that is taken to be representative of the sample as a whole; in essence, any stacking measurement makes this assumption implicitly.

With the assumption of two populations of uniform flux sources, the deblending equations above become an overdetermined (N+1)x2 linear system of the form
\begin{equation}
A x = b
\end{equation}
where x=$\{I_0, I_1\}$ is the vector of unknown, true fluxes of the (assumed uniform) target and non-target neighbors respectively, and $b$ is formally given by $b=\{f_i - n_i\}$, the $N+1$ dimensional vector of noiseless fluxes at the position of the target, $i=0$, and N nearby neighbors, $i\in\{1...N\}$.  In practice, $b = \{f_i\}$, the measured fluxes at these positions, as the noise is unknown with $\langle n_i \rangle=0$.  The coefficients of the (N+1)$\times$2 dimensional matrix $A$ are given by 
\begin{equation}
A_{i0} = \sum_{k=0}^{N} \delta_{k0} e^{-r_{ik}^2 / 2\Sigma^2} 
\end{equation}
\begin{equation}
A_{i1} = \sum_{k=0}^{N} \delta_{k1} e^{-r_{ik}^2 / 2\Sigma^2}
\end{equation}
where $\delta_{k0} = 1$ if source $k$ is a stacking target ($k=0$), and $\delta_{k0} = 0$ if source $k$ is not a stacking target, and $\delta_{k1} = 1$ if source k is a non-target neighbor ($k\in\{1...N\}$) and $\delta_{k1} = 0$ if source $k$ is a stacking target.  This system is overdetermined for $N>1$, and can be solved in a least squares sense using singular value decomposition.  The method is easily generalized to a larger number of non-target source categories.  In simulations below, we refer to this method as co-deblending.

\subsection{Global deblending}
\label{GlobalDeblendingSection}

We now introduce a generalization of the method of deblending discussed in Section \ref{CoDeblendingSection} that we expect to be more robust in the presence of noise than other methods.  As the number of neighbors, N, to a given target position is increased to be the size of the entire catalog of known objects, $\Lambda$, the method of co-deblending conceptually becomes a method of stacking and deblending simultaneously.  By adding together the deblend equations that result from considering each of the target stacking positions, and similarly combining the equations for the non-target positions, one obtains a system of equations that is exactly solvable for the blending corrected, averaged intrinsic flux of target and non-target populations.  

In this case, the measured flux values at stacking target positions are denoted by $f_{0j}$, for clarity, to distinguish them from the measured fluxes at non-target positions, denoted as $f_{1j}$.  The measured flux at each object position is the sum of contributions from all known object positions, whether stacking targets or non-targets, attenuated by their respective Gaussian beamwidth factors, $\alpha_{kj}$.\footnote{in implementing this algorithm,  the beamwidth factors for object separations that are many multiples of the beamwidth are set to zero to prevent accumulating round off errors.}  This procedure accounts for all neighbors of each source rather than a fixed number.

The measured fluxes at the target positions, $f_{0j}$, in the presence of noise, $n_{0j}$, are related to the underlying true object fluxes by
\begin{equation}
n_{0j}+ \sum_{k=0}^{\Lambda-1} \delta_{k0} \alpha_{kj}I_0 + \sum_{k=0}^{\Lambda-1} \delta_{k1} \alpha_{kj}I_1  = f_{0j}
\label{EQ:gdbtarget}
\end{equation}
where $\delta_{k0} = 1$ if source k is a stacking target and $\delta_{k0} = 0$ if source k is non-target and $\delta_{k1} = 1$ if source k is a non-target and $\delta_{k1} = 0$ if source k is a stacking target.  The sum is over all $\Lambda$ object positions, including target and non-target positions.  Similarly, the measured flux at the non-target positions, $f_{1j}$ in the presence of noise $n_{1j}$ is related to the true object fluxes by
\begin{equation}
n_{1j}+ \sum_{k=0}^{\Lambda-1} \delta_{k0} \alpha_{kj}I_0 + \sum_{k=0}^{\Lambda-1} \delta_{k1} \alpha_{kj}I_1  =  f_{1j}
\label{EQ:gdbnontarget}
\end{equation}
where the sum is over all $\Lambda$ object positions.

Adding together Equations \ref{EQ:gdbtarget} and \ref{EQ:gdbnontarget} for all target and non-target objects respectively yields an aggregate system of two equations and two unknowns that can be solved exactly for the unknown, average true object fluxes, $I_0$ and $I_1$, where the stacking target flux $I_0$ is principally of interest.

For the stacking target objects, Equation \ref{EQ:gdbtarget} is summed over all $R$ target positions:
\begin{equation}
 \sum_{j=0}^{R-1}n_{0j}+ \sum_{j=0}^{R-1} \sum_{k=0}^{\Lambda-1} \delta_{k0} \alpha_{kj}I_0 + \sum_{j=0}^{R-1} \sum_{k=0}^{\Lambda-1} \delta_{k1} \alpha_{kj}I_1 = \sum_{j=0}^{R-1} f_{0j} \label{eq:f}
\end{equation}
For the non-target objects, Equation \ref{EQ:gdbnontarget} is summed over all  $F \equiv \Lambda - R$ non-target positions:
\begin{equation}
 \sum_{j=0}^{F-1}n_{1j}+ \sum_{j=0}^{F-1} \sum_{k=0}^{\Lambda-1} \delta_{k0} \alpha_{kj}I_0 + \sum_{j=0}^{F-1} \sum_{k=0}^{\Lambda-1} \delta_{k1} \alpha_{kj}I_1 = \sum_{j=0}^{F-1} f_{1j} \label{eq:g}
\end{equation}
In the absence of noise, Equations~\ref{eq:f} and \ref{eq:g} constitute a 2x2 system of the form Ax = b that can be solved exactly for the vector of average true object fluxes, x$\equiv \{I_0, I_1\}$, where x = $A^{-1} b$.  This method can be generalized to include more object types $f_{2j}$, $f_{3j}$, ... in which case the linear system to be solved becomes 3x3, 4x4 etc.  Thus this method can be used to stack and deblend a catalog consisting of positions of multiple object types simultaneously.

In the presence of noise, simultaneous stacking and deblending is more robust than other methods, because the summations in Equations \ref{eq:f} and \ref{eq:g} effectively reduce the noise:   the sum of independent Gaussian noise terms tends to zero in the large sample limit.  Thus matrix multiplication yields the true flux estimates.

This method of stacking and deblending explicitly assumes that the flux of undetected objects in the stack are the same; as mentioned previously this assumption is implicit in any stacking methodology.  In the case of the aggregate flux being concentrated in just a few target objects, then an individual detection for those objects would likely have resulted and stacking would not normally have been applied to such a sample.  Correct interpretation of the results of any stacking algorithm demands recognizing the caveat that stacking merely estimates the average flux of the sample of undetected objects and does not reproduce the entire flux distribution of this sample.  However, the global deblending algorithm easily incorporates multiple sub-samples into the algorithm, which can be applied to the case where flux is suspected to be distributed non-uniformly among the sample of target objects.   

Conversely, there may be sources present in the data that are not present in the target list or catalog of non-target objects.  If these sources are spatially uncorrelated with the target objects, then these unaccounted-for sources of confusion will add scatter but not bias to the flux estimate.  It is important to use a sufficiently inclusive catalog that contains most or all of the objects that are spatially correlated with the target objects. 

\section{Simulations of deblending}
\label{SimulationsSection}

To validate the different approaches to stacking and deblending presented in Section \ref{MethodologySection}, we simulated artificial sources added to the LESS residual (source-subtracted) sub-millimeter image, which provides realistic noise.  We implemented several stacking and deblending algorithms and tested them on these artificial data.   

Sources were modeled in the data as having a Gaussian PSF with width equal to the LABOCA instrument beamwidth.   Flux densities were chosen to be either uniform among all sources, or drawn from a truncated power law distribution.  This latter distribution simulated the actual submillimeter source distribution in the ECDF-S, and is discussed in detail below.  The actual flux densities were chosen such that the signal to noise ratio of the final, stacked signal would be approximately between 1-100, the range for a plausible, stacking detection in real data.  

The spatial distribution of sources in the simulations were chosen to be identical to the positions of $K$ selected galaxies from the MUSYC ECDF-S survey.  Target positions were chosen to correspond to sBzK galaxies; non-target positions were drawn from the remaining $K<20$ galaxy catalog.  Thus the simulated source positions accurately reproduced the clustering statistics of real galaxies.  A subset of the full ECDF-S catalog data consisting of objects within a 0.7\degrees $\times$0.2\degrees~region centered in the LABOCA map were used such that 111 target objects were stacked in each simulation, and each simulated data set contained 1534 known object positions in total.  The 0.14 square degree subregion of the simulated data and noise for one random realization are illustrated in the right panel of Figure \ref{FIG:ECDFSData}.

Noise realizations were drawn from randomly chosen subsections of the LESS, bright source (SNR $>3.7 \sigma$) subtracted, residual map.  The distribution of residuals was approximately Gaussian with zero mean and 1$\sigma$ width $\approx$ 1.3 mJy, however there was a significant positive excess due to faint, individually undetected sources.  Each noise realization consisted of the LESS residual map translated and wrapped around by a uniform, random distance in RA (0-0.7\degrees) and Dec (0-0.2\degrees).  Artificial sources were added to the noise realization prior to stacking.

Monte Carlo simulations consisted of stacking and deblending the target source positions to estimate the weighted average, stacked target flux and signal to noise ratio for each realization, according to the various stacking algorithms.  The algorithms consisted of stacking with no deblending, standard deblending with 1, 2 or more nearby neighbors, co-deblending with 1, 2 or more nearby neighbors, and simultaneous stacking and deblending, referred to as global deblending.  

Each simulation consisted of 10$^4$ realizations of artificial sources and noise.  For each realization, the estimated, stacked target flux was computed according to each algorithm.  These estimated fluxes were compared to the true source fluxes to compute the rms error of each stacking estimate within the set of realizations of the simulation.  Comparing this RMS error to the estimated, stacked signal to noise ratio allowed for evaluation of the various algorithms.  For example, an ideal stacking signal to noise ratio of 1.00 would yield an rms error of 100\%, because the signal is equal in amplitude to the noise.  The RMS errors of the resulting stacking detections were computed for each set of random realizations, at each source flux density.  In these simulations, the individual source SNR was very low,\footnote{e.g. SNR$_{indiv}$ $\approx SNR_{stack} /  \sqrt{111}$ for stacking 111 individual sources} in accordance with the premise of stacking -- very low individual SNR measurements were combined to yield a significant aggregate detection.  

We have performed additional simulations with artificial data sets to validate the deblending algorithms for various source configurations with Gaussian and real noise maps.  We performed simulations with a grid of artificial sources, each source having a single blended neighbor.  As the neighbor separation was increased, thereby reducing overlap between sources, the stacking estimates all converged to the same, true result.  We also verified that as the flux ratio between stacking target and blended neighbor was diminished, such that the blended neighbor was much fainter than the target, the stacking estimates converged to the same, true result.  Finally, in simulations with the real distribution of galaxy positions, we verified that as non-target sources were removed at random from the data, making the data more sparsely distributed, and therefore blending less important, all stacking estimates converged to the same, true result.  These tests gave confidence that the stacking algorithms were correctly implemented. 

\section{Results}

\begin{figure}[h]
\begin{center}

\includegraphics[scale=0.40]{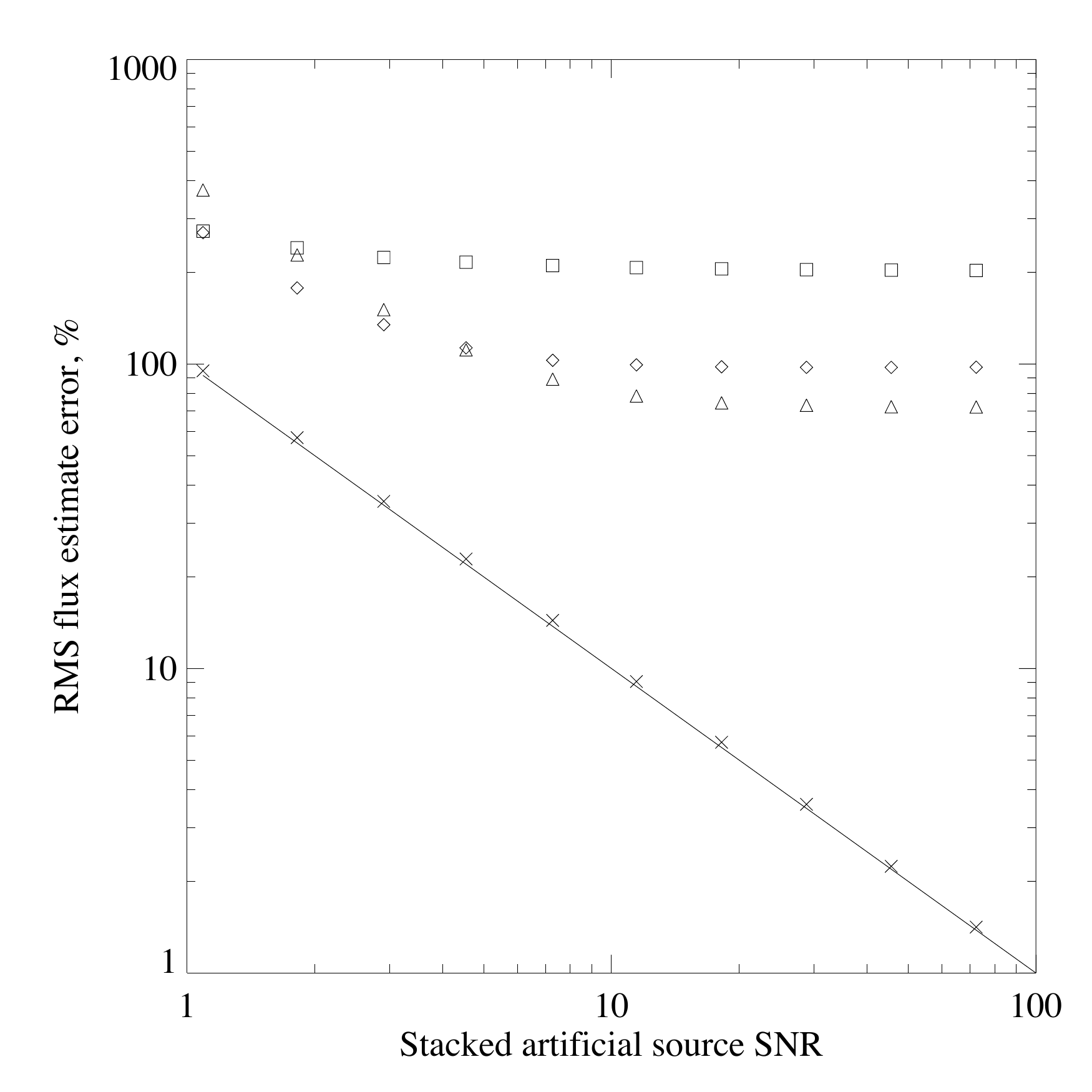}
\includegraphics[scale=0.40]{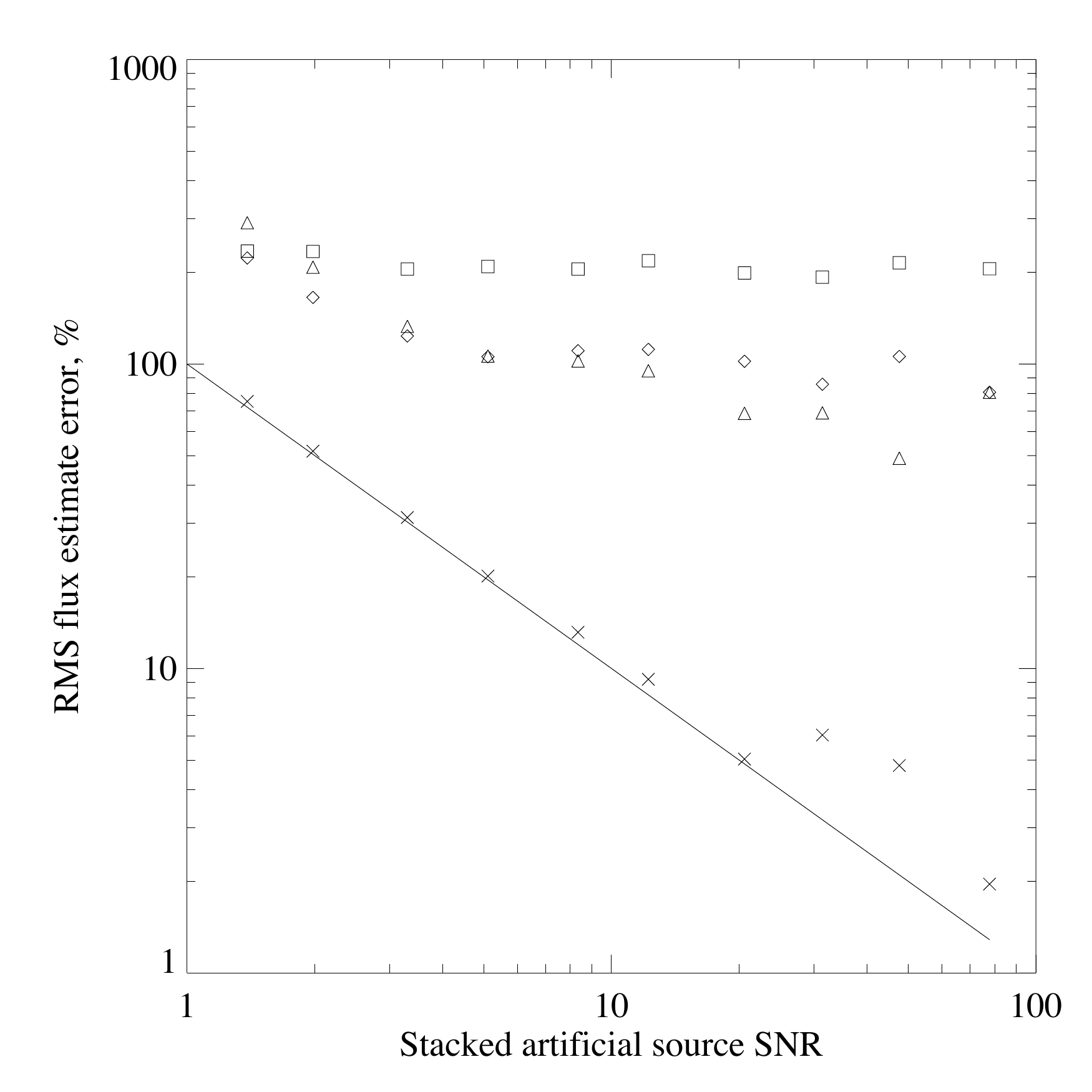}

\vspace{-0.5cm}
\end{center}
\caption{\footnotesize
{Results of simulations of stacking and deblending algorithms.  RMS flux estimate error vs. artificial source signal to noise ratio.  Left panel:  source flux densities were uniform.  Right panel:  source flux densities were drawn from a power law distribution.  Results of 11 simulations are depicted, with stacked SNR between 1-100.  Deblending algorithms included nearest and next nearest neighbors.  Squares - no deblending.  Triangles - standard deblending.  Diamonds - co-deblending.  Crosses - global deblending.  Solid line represents an ideal stacking algorithm.}}

\label{FIG:RMSErrorVSStackedSNR}
\end{figure}

The main results of simulations are illustrated in Figure \ref{FIG:RMSErrorVSStackedSNR}, which plots RMS error of the stacking estimate for each set of realizations vs. the true stacked SNR.  The horizontal axis of each figure refers to the signal to noise ratio of the stacking detection, assuming no systematics, i.e. Poisson statistics. 

\begin{figure}[h]
\begin{center}

\includegraphics[scale=0.40]{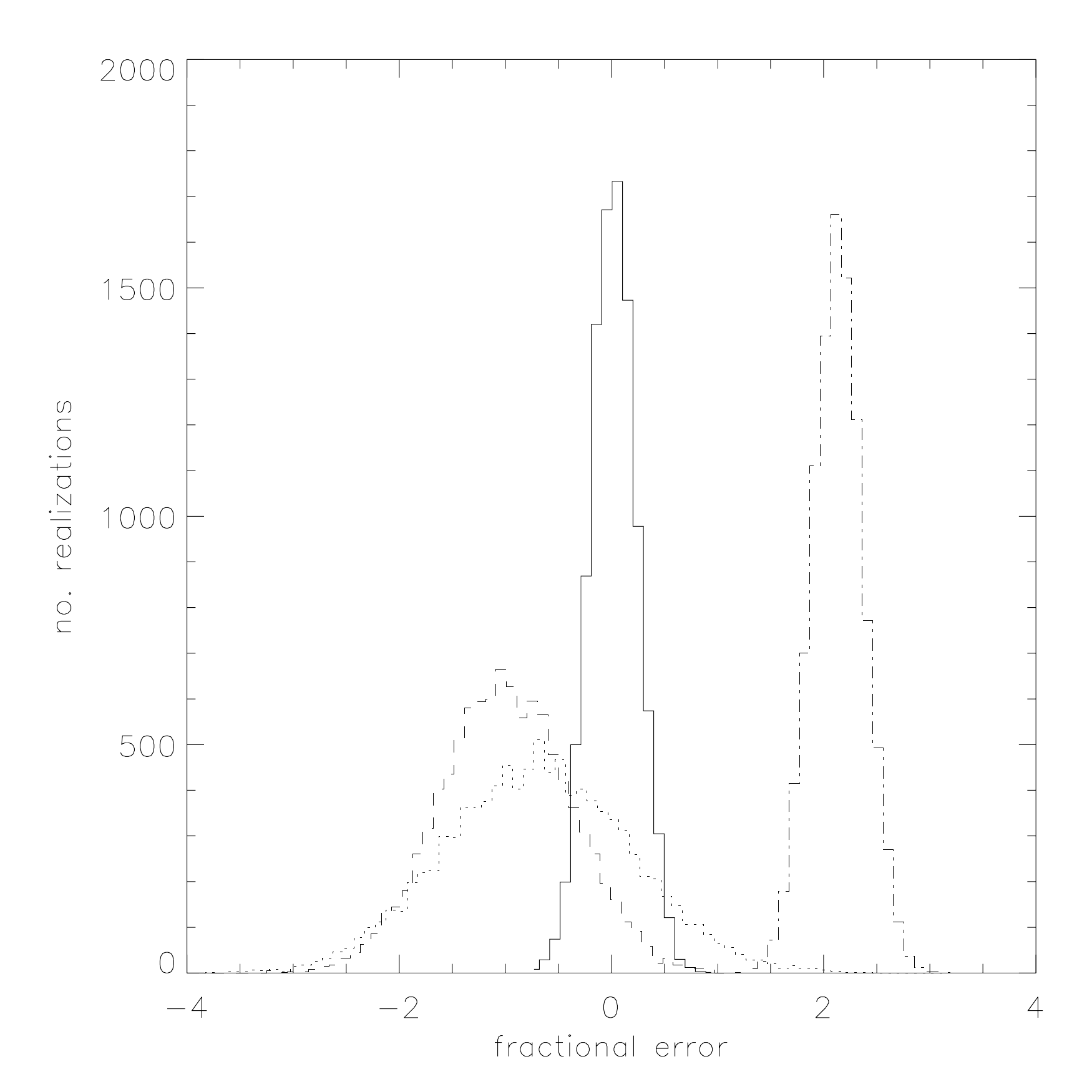}
\includegraphics[scale=0.40]{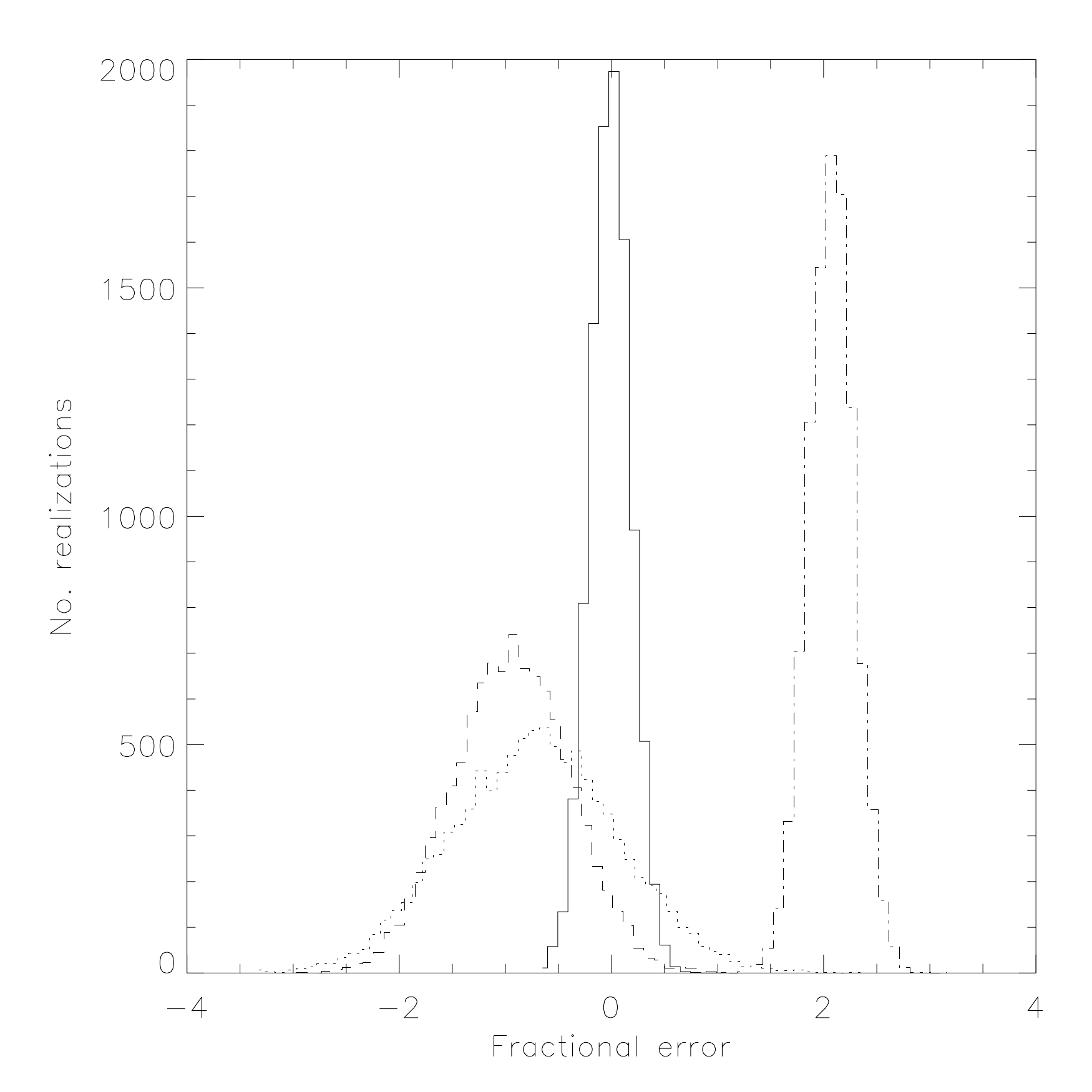}

\vspace{-0.5cm}
\end{center}
\caption{\footnotesize
{Histograms of stacking signal errors found from simulated SNR $\approx$ 5 stacking detections.  Left panel:  source flux densities were uniform.  Right panel:  source flux densities were drawn from a power law distribution.  Solid curve - global deblending, dotted - standard deblending, dash - co-deblending, dash-dot - no deblending.  Global deblending exhibits the smallest bias of the deblending estimates, and shares the smallest scatter with no deblending.}}

\label{FIG:DeblendHistograms}
\end{figure}

\begin{figure}[h]
\begin{center}

\includegraphics[scale=0.40]{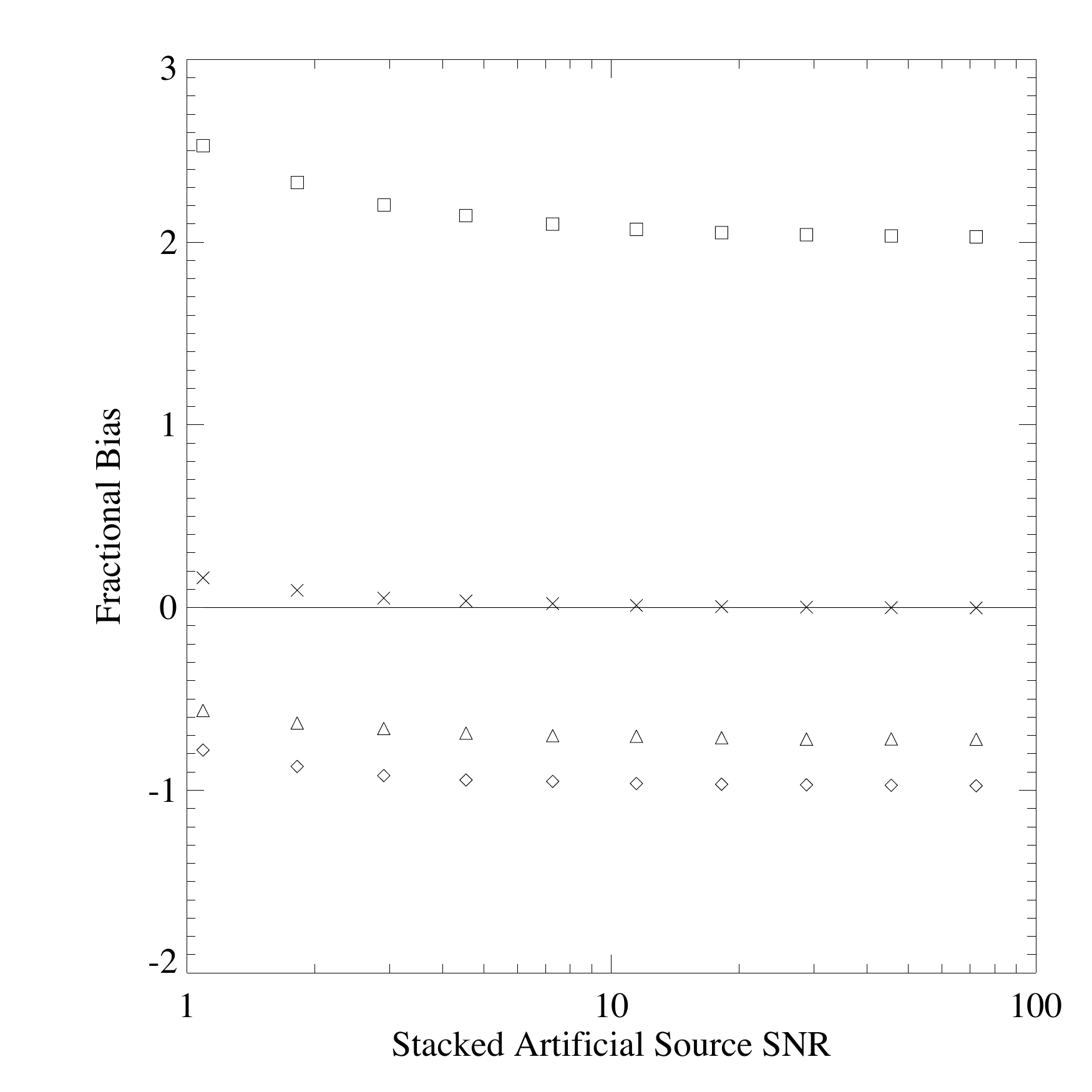}
\includegraphics[scale=0.40]{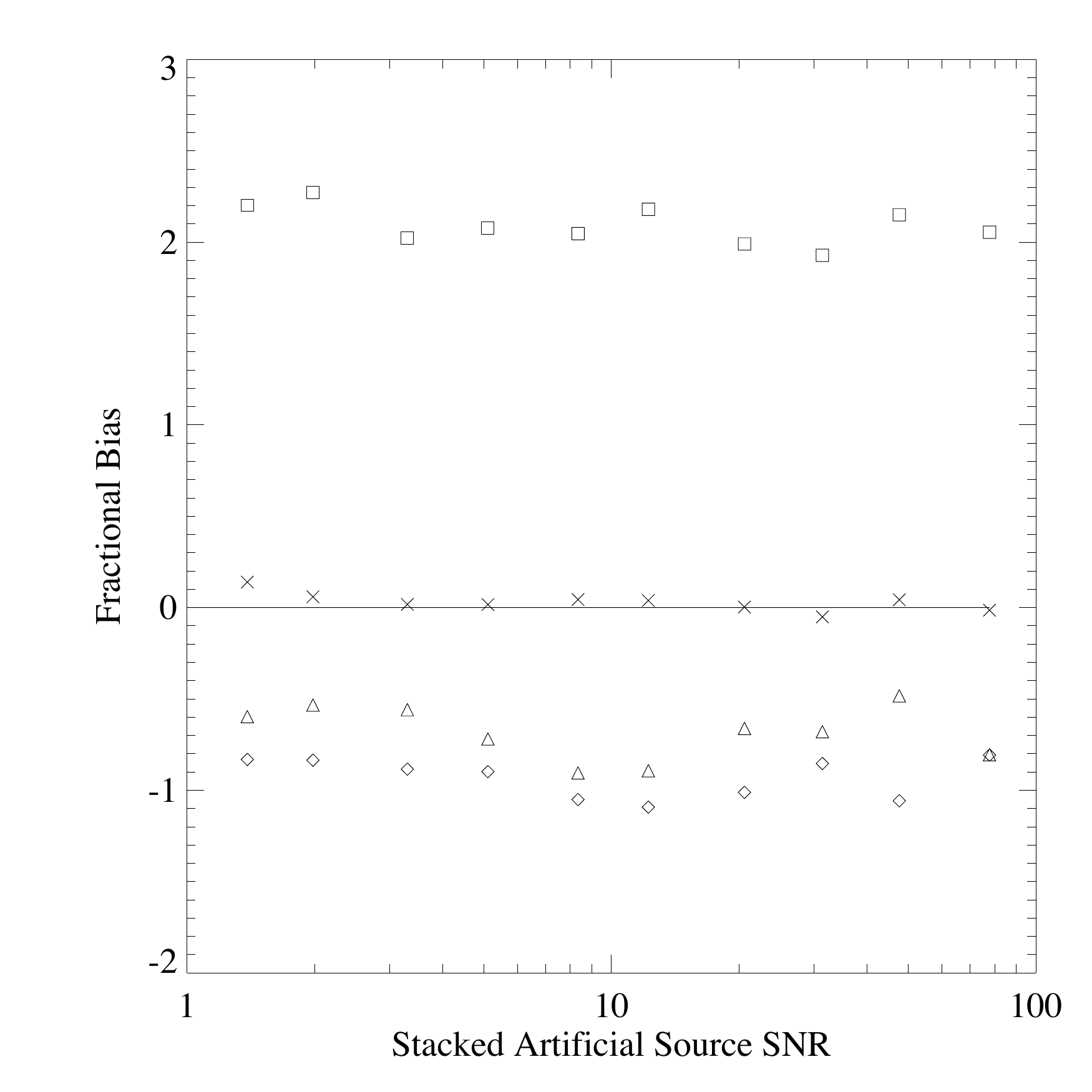}

\vspace{-0.5cm}
\end{center}
\caption{\footnotesize
{Bias vs. SNR for the simulations of Figure \ref{FIG:RMSErrorVSStackedSNR}.  Left panel:  source flux densities were uniform.  Right panel:  source flux densities were drawn from a power law distribution.  Squares - no deblending.  Triangles - standard deblending.  Diamonds - co-deblending.  Crosses - global deblending.  The horizontal line indicates zero bias and is shown for clarity.}}

\label{FIG:BiasVsSNR}
\end{figure}

\begin{figure}[h]
\begin{center}

\includegraphics[scale=0.40]{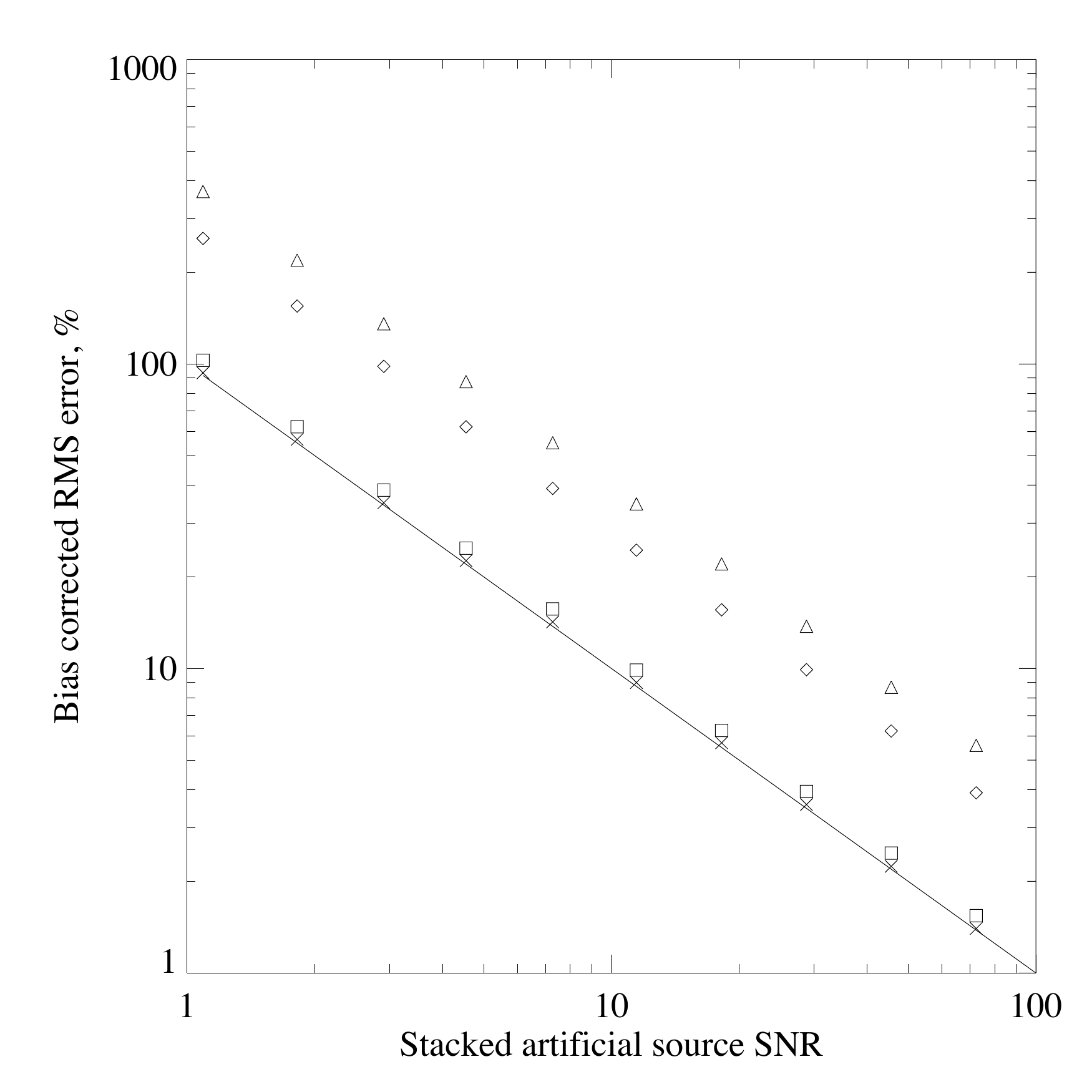}
\includegraphics[scale=0.40]{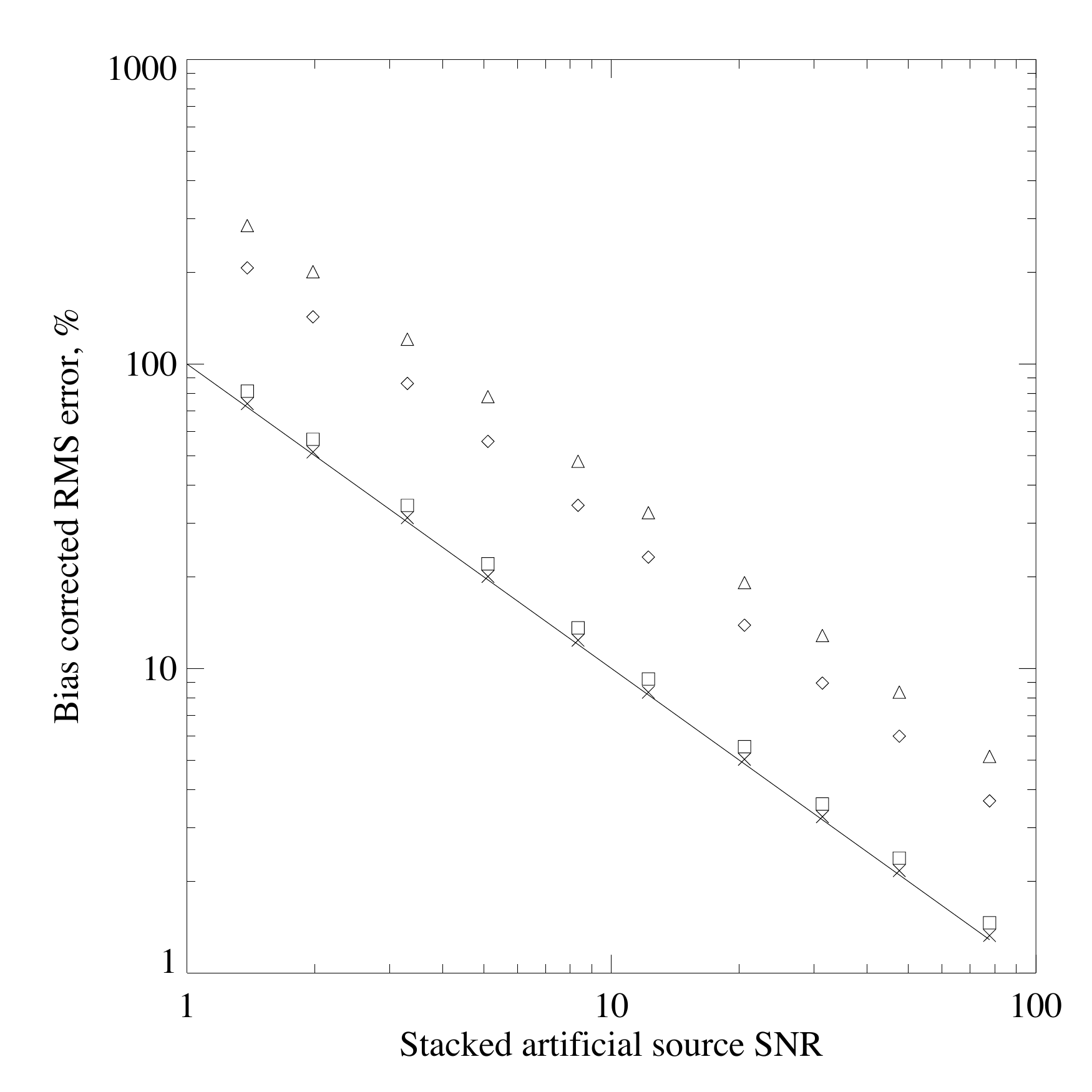}

\vspace{-0.5cm}
\end{center}
\caption{\footnotesize
{Bias corrected error distributions vs. SNR for the simulations of Figure \ref{FIG:RMSErrorVSStackedSNR}.  Left panel:   source flux densities were uniform.  Right panel:  source flux densities were drawn from a power law distribution.  Squares - no deblending.  Triangles - standard deblending.  Diamonds - co-deblending.  Crosses - global deblending.  Solid line represents an ideal (bias corrected) stacking algorithm.}}

\label{FIG:BiasCorrRMSvsSNR}
\end{figure}

In Figures  \ref{FIG:RMSErrorVSStackedSNR} - \ref{FIG:BiasCorrRMSvsSNR}, the left panel illustrates the results obtained with uniform flux density sources, and the right panel corresponds to sources with flux densities drawn from a power law distribution as follows:  the power law index was -3.2, chosen to simulate the submillimeter source distribution observed in the ECDF-S, and reported in \citet{2009arXiv0910.2821W}; the distribution was
truncated above such that all sources were below the LABOCA detection threshold.  We assumed that the 8000 K$<$20 MUSYC sources comprised the brightest sub-threshold sub-millimeter sources according to the indicated power law distribution and used this integral constraint to solve for the lower limit of the power law distribution, which was $\approx$0.5 mJy. This procedure yielded a stacking detection with SNR $\approx$10. The flux values were scaled by a uniform constant in each simulation so that the final stacked detection spanned a range of signal to noise ratios between 1 and 100.  

Previous stacking analysis of the submillimeter flux in the ECDF-S suggested that K$<$20 sources contribute 15\% of the submillimeter background \citep{2009arXiv0904.0028G}.  All fluctuations due to sources that comprise the submillimeter background are included in the LABOCA residual map, and hence are accounted for in these simulations, although the fluxes of individual undetected sources are by definition unknown.  The deliberate addition of faint artificial sources serves to make the simulations somewhat more noisy than the actual data and the simulation results correspondingly somewhat more conservative.  However, these artificial sources test the effectiveness of the stacking algorithms at deblending targets from other populations that may have spatial correlations with them, which is precisely the difficulty encountered in practice.

Figure \ref{FIG:RMSErrorVSStackedSNR} illustrates the importance of deblending.  An ideal stacking algorithm would achieve statistical RMS errors indicated by the solid line in the figure.  Without deblending (square symbols), significant additional systematic errors occur at all signal to noise ratios due to confusion.  Global deblending (cross symbols) performs best, with RMS errors close to the statistical limit through the entire range of SNR.  The RMS error from global deblending is reduced by more than a factor of three compared to other algorithms at all signal to noise ratios and by an order of magnitude or more at high signal to noise ratios (SNR $>$ 10).

The observed bias in the estimates from standard and co-deblending is not caused by correlations or other particular noise features of the LABOCA residual map; simulations with independent, Gaussian noise at each pixel exhibited similar bias.  However, the bias was reduced by $\sim 25\%$ in simulations where the number of objects in the stack was tripled.  This result  suggests that the bias originates from averaging of quantities in the stacking procedure.

The other deblending algorithms used nearest and next-nearest neighbors in the deblending calculations.  Varying the number of neighbors used in deblending beyond N=3 generally worsened the results for deblending and co-deblending algorithms; this observation is not surprising given the discussion in section \ref{GlobalDeblendingSection} about matrix inversion of noisy data.

Figure \ref{FIG:DeblendHistograms} illustrates the distributions of fractional errors from a representative simulation.  Histograms of the stacking estimate error for a typical simulation, with stacked, artificial SNR $\approx$5 are illustrated for the no deblending, standard deblending, co-deblending and global deblending methods.  The mean value of each distribution indicates the amount of bias in the flux estimate, and the width of the distribution indicates its variance.  These distributions indicate that stacking without deblending yields the most significantly biased flux estimate, however it also has low variance.  Thus, stacking without deblending does nothing to introduce additional noise to the data; it simply fails to subtract the contamination by neighbors.  Standard deblending and co-deblending have reduced, but still significant bias and they add scatter to the distributions.  

The error estimates from the simulations of Figure \ref{FIG:DeblendHistograms} are compared to each other and the ideal minimum variance statistical errors in Table 1.  Column 1 indicates each of the 4 stacking and deblending algorithms under consideration.  Columns 2 (5) indicate the statistical errors corresponding to a stacked, weighted average RMS from the LABOCA data for sources with uniform flux (power law flux distribution); they are the minimum values obtainable for estimated errors of an ideal stacking algorithm.  RMS errors for each pixel were obtained from an RMS map accompanying the LABOCA data, and the error for each stack was obtained from the usual formula for the weighted average.  Columns 3 (6) indicate actual errors,  defined as the RMS of the difference between the stacked and deblended flux estimate and the actual weighted average flux of the stacking targets, averaged over all realizations of the simulation, for sources with uniform flux (power law flux distribution).  Finally Columns 4 (7) indicate the bias corrected RMS error for sources with uniform flux (power law flux distribution).  Global deblending approaches the statistical minimum error for these simulations, as does bias corrected no deblending, whereas other algorithms have significantly larger errors.  

Global deblending greatly reduces bias while maintaining the minimum variance of the estimate.  The bias evident from the distributions in Figure \ref{FIG:DeblendHistograms} is illustrated for each SNR in Figure \ref{FIG:BiasVsSNR}.  At all SNR values, no deblending has the largest bias, standard deblending and co-deblending show a smaller bias, and global deblending shows the least bias.  

The bias corrected estimates are illustrated in Figure \ref{FIG:BiasCorrRMSvsSNR}.  The figure illustrates that bias correcting the no deblending estimate would yield comparable performance to global deblending.  In practice, global deblending would be the best way to estimate this bias, so it is the preferred approach.  The standard deblending and co-deblending algorithms cannot be simply corrected to the statistical limit because of the larger scatter in the distributions, evident from Figure \ref{FIG:DeblendHistograms}.  This scatter results from the difficulty of inverting a matrix of noisy data, as discussed in Section \ref{GlobalDeblendingSection}.  Confusion functions to add bias to the stacking estimate.  Figures \ref{FIG:BiasVsSNR} and \ref{FIG:BiasCorrRMSvsSNR} indicate that standard deblending algorithms partially remove this bias, but at the expense of introducing additional scatter that cannot be corrected.  Global deblending significantly reduces bias and approaches minimum variance by summing over noisy data as much as possible before performing matrix inversion. 

\section{Conclusion}

The simultaneous stacking and deblending algorithm presented here demonstrated more than a factor of three improvement in RMS error over other deblending algorithms at all signal to noise ratios, and by an order of magnitude at large signal to noise ratios.  This improvement translates into a greater sensitivity for stacking analyses that will be applicable to present and future far infrared and submillimeter surveys.  For example, the Herschel SPIRE instrument will reach 1$\sigma$ survey depths of 3 mJy in 10-16 hours  at 18-36$''$ spatial resolution \citep{2008SPIE.7010E...4G}.  At this depth, and with 1,000 known source positions, stacking analyses can reach a depth of 0.1 mJy when limited only by statistical error.  This depth would be sufficient to detect stacking signals from color selected galaxies such as BzKs, DRGs and EROs that have been detected in stacking at 870 $\mu$m with fluxes in the 0.2-0.4 mJy range with the ground based LABOCA instrument \citep{2009arXiv0904.0028G}.  However, with the factor of three poorer sensitivity caused by standard deblending algorithms, these objects would remain out of reach to stacking analysis with Herschel, or their reported stacked fluxes would be significantly biased.

\acknowledgments

Support for this work was provided by the National Science Foundation under grant AST-0807570, by the Department of Energy under grant DE-FG02-08ER41561, and by NASA through an award issued by JPL/Caltech.  We acknowledge valuable conversations and comments on this manuscript by Nicholas A. Bond, Fabian Walter, Thomas R. Greve, Felipe Menanteau, Ian Smail, and Axel Weiss, and we thank LESS for providing the residual map used for simulations, obtained from LABOCA APEX RUN IDs: 078.F-9028(A), 079.F-9500(A), 080.A-3023(A), and 081.F-9500(A)

\bibliographystyle{apj}                       


\bibliography{apj-jour,StackingBibliography}

\begin{deluxetable}{rccccccc} 
\tablecolumns{8} 
\tablewidth{0pc} 
\tablecaption{Comparison of Error Estimation} 
\tablehead{ 
\colhead{}    &  \multicolumn{3}{c}{Uniform Source Flux Distribution} &   \colhead{}   & 
\multicolumn{3}{c}{Power Law Source Flux Distribution} \\ 
\cline{2-4} \cline{6-8} \\ 
\colhead{Deblend} & \colhead{Statistical}   & \colhead{Actual}    & \colhead{Bias Corr.} & 
\colhead{}    & \colhead{Statistical}   & \colhead{Actual}    & \colhead{Bias Corr.} \\
\colhead{Method} & \colhead{RMS (mJy)}   & \colhead{RMS (mJy)}    & \colhead{RMS (mJy)} & 
\colhead{}    & \colhead{RMS (mJy)}   & \colhead{RMS (mJy)}    & \colhead{RMS (mJy)} }

\startdata 
None & 0.11 &1.08 & 0.13 && 0.11 & 1.11 & 0.12 \\
Standard & 0.11 & 0.56 & 0.44 && 0.11 & 0.61 & 0.44 \\
Co-deblend& 0.11 & 0.56 & 0.31 && 0.11 & 0.59 & 0.31 \\
Global & 0.11 & 0.12 & 0.11 && 0.11 & 0.12 & 0.11 \\
\enddata 
\tablecomments{For simulated stacked flux of 0.50(0.55) $\pm$ 0.11 mJy for uniform (power law) source flux distribution (SNR $\approx$ 5 stacking detection)}

\end{deluxetable} 
\end{document}